\documentclass[12pt]{iopart}

\usepackage{graphicx}
\begin{document}

\title{Ultra accurate personalized recommendation via eliminating redundant correlations}

\author{Tao Zhou$^{1,2}$, Ri-Qi Su$^1$, Run-Ran Liu$^1$, Luo-Luo Jiang$^1$, Bing-Hong Wang$^{1,3}$, Yi-Cheng Zhang$^{2,3}$}

\address{$^1$Department of Modern Physics and Nonlinear Science Center,
University of Science and Technology of China, Hefei Anhui 230026,
P. R. China \\
$^2$ Department of Physics, University of Fribourg, Chemin du Muse
3, CH-1700 Fribourg, Switzerland\\
$^3$Research Center for Complex System Science, University of
Shanghai for Science and Technology, Shanghai 200093, P. R. China}

\ead{bhwang@ustc.edu.cn}

\begin{abstract}
In this paper, based on a weighted projection of bipartite
user-object network, we introduce a personalized recommendation
algorithm, called the \emph{network-based inference} (NBI), which
has higher accuracy than the classical algorithm, namely
\emph{collaborative filtering}. In the NBI, the correlation
resulting from a specific attribute may be repeatedly counted in the
cumulative recommendations from different objects. By considering
the higher order correlations, we design an improved algorithm that
can, to some extent, eliminate the redundant correlations. We test
our algorithm on two benchmark data sets, \emph{MovieLens} and
\emph{Netflix}. Compared with the NBI, the algorithmic accuracy,
measured by the ranking score, can be further improved by 23\% for
\emph{MovieLens} and 22\% for \emph{Netflix}, respectively. The
present algorithm can even outperform the \emph{Latent Dirichlet
Allocation} algorithm, which requires much longer computational
time. Furthermore, most of the previous studies considered the
algorithmic accuracy only, in this paper, we argue that the
diversity and popularity, as two significant criteria of algorithmic
performance, should also be taken into account. With more or less
the same accuracy, an algorithm giving higher diversity and lower
popularity is more favorable. Numerical results show that the
present algorithm can outperform the standard one simultaneously in
all five adopted metrics: lower ranking score and higher precision
for accuracy, larger Hamming distance and lower intra-similarity for
diversity, as well as smaller average degree for popularity.
\end{abstract}

\maketitle

\section{Introduction}
The exponential growth of the Internet \cite{Zhang2008} and World
Wide Web \cite{Broder2000} confronts people with an information
overload: They face too much data and sources to be able to find out
those most relevant for them. Indeed, people may choose from
thousands of movies, millions of books, and billions of web pages.
The amount of information is increasing more quickly than our
processing ability, thus evaluating all these alternatives and then
making choice becomes infeasible. A landmark for information
filtering is the use of search engine \cite{Brin1998,Kleinberg1999},
by which users could find the relevant webpages with the help of
properly chosen tags. However, the search engine has two essential
disadvantages. On the one hand, it does not take into account
personalization and thus returns the same results for people with
far different habits. So, if a user's habits are different from the
mainstream, even with some ``right tags", it is hard for him to find
out what he likes from the countless searching results. On the other
hand, some tastes, such as the feelings of music and poem, can not
be expressed by tags, even language. The search engine, based on tag
matching, will lose its effectiveness in those cases.

Thus far, the most promising way to efficiently filter out the
information overload is to provide personalized recommendations.
That is to say, using the personal information of a user (i.e., the
historical track of this user's activities) to uncover his habits
and to consider them in the recommendation. For example,
\emph{Amazon.com} uses one's purchase record to recommend books
\cite{Linden2003}, \emph{AdaptiveInfo.com} uses one's reading
history to recommend news \cite{Billsus2002}, and the \emph{TiVo}
digital video system recommends TV shows and movies on the basis of
users' viewing patterns and ratings \cite{Ali2004}.

Motivated by the significance in economy and society
\cite{Schafer2001}, the design of an efficient recommendation
algorithm becomes a joint focus from engineering science to
marketing practice, from mathematical analysis to physics community
(see the review article \cite{Adomavicius2005} and the references
therein). Various kinds of algorithms have been proposed, including
collaborative filtering \cite{Herlocker2004}, content-based analysis
\cite{Pazzani2007}, spectral analysis \cite{Maslov2001}, iteratively
self-consistent refinement \cite{Ren2008}, principle component
analysis \cite{Goldberg2001}, and so on.

Very recently some physical dynamics, including heat conduction
process \cite{Zhang2007a} and mass/energy diffusion
\cite{Zhou2007,Zhang2007b,Zhou2008}, have found applications in
personalized recommendation. These physical approaches have been
demonstrated to be both highly efficient and of low computational
complexity. In this paper, we will first introduce a network-based
recommendation algorithm, called the \emph{network-based inference}
(NBI) \cite{Zhou2007}, which has higher accuracy than the classical
algorithm, namely collaborative filtering. In the NBI, the
correlation resulting from a specific attribute may be repeatedly
counted in the cumulative recommendations from different objects. By
considering the higher order correlations, we next design a higher
effective algorithm that can, to some extent, eliminate the
redundant correlations. Numerical results show that, the improved
algorithm has much higher accuracy and can provide more diverse and
less popular recommendations.

\section{Network-Based Inference for Personal Recommendation}
A recommendation system consists of users and objects, and each user
has collected some objects. Denoting the object-set as
$O=\{o_1,o_2,\cdots,o_n\}$ and user-set as
$U=\{u_1,u_2,\cdots,u_m\}$, the recommendation system can be fully
described by a bipartite network with $n+m$ nodes, where an object
is connected with a user if and only if this object has been
collected by this user. Connections between two users or two objects
are not allowed. Based on the bipartite user-object network, an
object-object network can be constructed, where each node represents
an object, and two objects are connected if and only if they have
been collected simultaneously by at least one user. We assume a
certain amount of resource (i.e., recommendation power) is
associated with each object, and the weight $w_{ij}$ represents the
proportion of the resource $o_j$ would like to distribute to $o_i$.
For example, in the book-selling system, the weight $w_{ij}$
contributes to the strength of recommending the book $o_i$ to a
customer provided he has already bought the book $o_j$.

\begin{figure}
\scalebox{1.5}[1.5]{\includegraphics{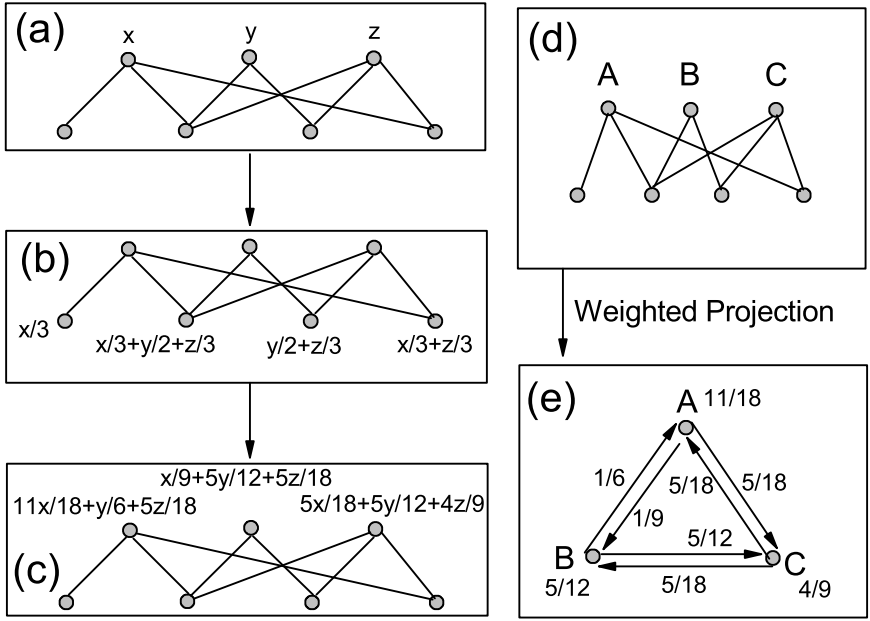}} \caption{Illustration
of the resource-allocation process in a bipartite network. In plots
(a), (b), and (c), the upper three are $X$-nodes, and the lower four
are $Y$-nodes. The whole process consists of two steps: First, the
resource flows from $X$ to $Y$ (a$\rightarrow$b), and then returns
to $X$ (b$\rightarrow$c). Indeed, the process from (a) to (c) can be
considered as a weighted projection of a bipartite network, shown as
(d)$\rightarrow$(e). The weight located on the directed edge
A$\rightarrow$B means the fraction of resource node A would transfer
to node B. The weights of self-connections are also labelled beside
the corresponding nodes.}
\end{figure}

The weight $w_{ij}$ can be determined following a network-based
resource-allocation process \cite{Ou2007} where each object
distributes its initial resource equally to all the users who have
collected it, and then each user sends back what he has received
equally to all the objects he has collected. Figure 1 gives a simple
example, where the three $X$-nodes are initially assigned weights
$x$, $y$ and $z$. The resource-allocation process consists of two
steps; first from $X$ to $Y$, then back to $X$. The amount of
resource after each step is marked in Fig. 1(b) and Fig. 1(c),
respectively. Merging these two steps into one, the final resource
located in the three $X$-nodes, denoted by $x'$, $y'$ and $z'$, can
be obtained as:
\begin{equation}
 \left(
     \begin{array}{c}
        x' \\
        y' \\
        z' \\
     \end{array}
     \right)
 =\left(
    \begin{array}{ccc}
        11/18 & 1/6 & 5/18 \\
        1/9 & 5/12 & 5/18 \\
        5/18 & 5/12 & 4/9
    \end{array}
    \right)
 \left(
     \begin{array}{c}
        x \\
        y \\
        z \\
     \end{array}
     \right).
\end{equation}
According to the above description, this $3\times 3$ matrix is the
very weighted matrix we want. Clearly, this weighted matrix,
equivalent to a weighted projection network of $X$-nodes, is
independent of the initial resources assigned to $X$-nodes. A
network representation is shown in Fig. 1(d) and 1(e). For a general
user-object network, the weighted projection onto object-object
network reads \cite{Zhou2007}:
\begin{equation}
w_{ij}=\frac{1}{k(o_j)}\sum^m_{l=1}\frac{a_{il}a_{jl}}{k(u_l)},
\end{equation}
where $k(o_j)=\sum^m_{i=1}a_{ji}$ and $k(u_l)=\sum^n_{i=1}a_{il}$
denote the degrees of object $o_j$ and user $u_l$, and \{$a_{il}$\}
is an $n\times m$ adjacent matrix of the bipartite user-object
network, defined as:
\begin{equation}
\label{cases}
a_{il}=\cases{1& $o_i$ is collected by $u_l$,\\
0&otherwise.\\}
\end{equation}

For a given user $u_i$, we assign some resource (i.e.,
recommendation power) on those objects already collected by $u_i$.
In the simplest case, the initial resource vector $\mathbf{f}$ can
be set as
\begin{equation}
f_j=a_{ji}.
\end{equation}
That is to say, if the object $o_j$ has been collected by $u_i$,
then its initial resource is unit, otherwise it is zero. After the
resource-allocation process, the final resource vector is
\begin{equation}
\mathbf{f}'=W\mathbf{f}.
\end{equation}
Accordingly, all $u_i$'s uncollected objects $o_j$ ($1\leq j \leq
n$, $a_{ji}=0$) are sorted in the descending order of $f'_j$, and
those objects with the highest values of final resource are
recommended. We call this method \emph{network-based inference}
(NBI), since it is based on the weighted object-object network
\cite{Zhou2007}.

For comparison, we briefly introduce two classical recommendation
algorithms. The first is the so-called \emph{global ranking method}
(GRM), which sorts all the objects in the descending order of degree
and recommends those with the highest degrees. The second is the
most widely applied recommendation algorithm, named
\emph{collaborative filtering} (CF) \cite{Herlocker2004}. This
algorithm is based on measuring the similarity between users or
objects. The most widely used similarity measure, also adopted in
this paper, is the so-called \emph{S{\o}rensen index} (i.e., the
cosine similarity) \cite{Sorensen1948}. For two users $u_i$ and
$u_j$, their cosine similarity is defined as (for more local
similarity indices as well as the comparison of them, see the Refs.
\cite{Liben-Nowell2007,Zhou2009}):
\begin{equation}
s_{ij}=\frac{1}{\sqrt{k(u_i)k(u_j)}}\sum^n_{l=1}a_{li}a_{lj}.
\end{equation}
For any user-object pair $u_i-o_j$, if $u_i$ has not yet collected
$o_j$ (i.e., $a_{ji}=0$), the predicted score, $v_{ij}$ (to what
extent $u_i$ likes $o_j$), is given as
\begin{equation}
v_{ij}=\frac{\sum^m_{l=1,l\neq i}s_{li}a_{jl}}{\sum^m_{l=1,l\neq
i}s_{li}}.
\end{equation}
For any user $u_i$, all the nonzero $v_{ij}$ with $a_{ji}=0$ are
sorted in a descending order, and those objects in the top are
recommended. This algorithm is based on the similarity between user
pairs, we therefore call it user-based collaborative filtering,
abbreviated as UCF. The main idea embedded in the UCF is that the
target user will be recommended the objects collected by the users
sharing similar tastes. Analogously, the recommendation list can be
obtained by object-based collaborative filtering (OCF), that is, the
target user will be recommended objects similar to the ones he
preferred in the past (see Refs. \cite{Sarwar2001,Liu2009} the
investigation of OCF algorithms as well as the comparison between
UCF and OCF). Using also the S{\o}rensen index, the similarity
between two objects, $o_i$ and $o_j$, can be written as:
\begin{equation}
s^o_{ij}=\frac{1}{\sqrt{k(o_i)k(o_j)}}\sum^m_{l=1}a_{il}a_{jl},
\end{equation}
where the superscript emphasizes that this measure is for object
similarity. The predicted score, to what extent $u_i$ likes $o_j$,
is given as:
\begin{equation}
v_{ij}=\frac{\sum^n_{l=1,l\neq i}s^o_{jl}a_{li}}{\sum^n_{l=1,l\neq
i}s^o_{jl}}.
\end{equation}

To test the algorithmic accuracy, we use two benchmark data sets,
namely \emph{MovieLens} (http://www.grouplens.org/) and
\emph{Netflix} (http://www.netflixprize.com/). The \emph{MovieLens}
data consists of 1682 movies (objects) and 943 users, and users vote
movies using discrete ratings 1-5. We therefore applied a
coarse-graining method \cite{Zhou2007,Zhou2008}: a movie has been
collected by a user if and only if the giving rating is at least 3
(i.e. the user at least likes this movie). The original data
contains $10^5$ ratings, 85.25\% of which are $\geq 3$, thus after
coarse gaining the data contains 85250 user-object pairs. The
\emph{Netflix} data is a random sampling of the whole records of
user activities in \emph{Netflix.com}, consisting of 10000 users,
6000 movies and 824802 links. Similar to the MovieLens data, only
the links with ratings no less than 3 are kept. To test the
recommendation algorithms, the data set is randomly divided into two
parts: The training set contains 90\% of the data, and the remaining
10\% of data constitutes the probe. The training set is treated as
known information, while no information in the probe set is allowed
to be used for recommendation.

\begin{table}
\caption{Algorithmic performance for \emph{MovieLens} data. The
precision, intra-similarity, diversity and popularity are
corresponding to $L=50$. Heter-NBI is an abbreviation of NBI with
heterogenous initial resource distribution, proposed in Ref.
\cite{Zhou2008}. RE-NBI is an abbreviation of redundant-eliminated
NBI, the very algorithm presented in this paper. The parameters in
Heter-NBI and RE-NBI are set as the ones corresponding to the lowest
ranking scores (for Heter-NBI \cite{Zhou2008},
$\beta_{\texttt{opt}}=-0.80$; for RE-NBI, $a_{\texttt{opt}}=-0.75$).
Each number presented in this table is obtained by averaging over
five runs, each of which has an independently random division of
training set and probe. }
\begin{center}
\begin{tabular} {ccccccc}
  \hline \hline
   Algorithms     & Ranking Score  &  Precision &  Intra-Similarity & Hamming Distance & Popularity  \\
   \hline
   GRM & 0.140 & 0.054 & 0.408 & 0.398 & 259  \\
   UCF & 0.127 & 0.065 & 0.395 & 0.549 & 246  \\
   OCF & 0.111 & 0.070 & 0.412 & 0.669 & 214  \\
   NBI & 0.106 & 0.071 & 0.355 & 0.617 & 233  \\
   Heter-NBI & 0.101 & 0.073 & 0.341 & 0.682 & 220  \\
   RE-NBI & 0.082 & 0.085 & 0.326 & 0.788 & 189  \\
   \hline \hline
    \end{tabular}
\end{center}
\end{table}

\begin{table}
\caption{Algorithmic performance for \emph{Netflix} data. The
precision, intra-similarity, Hamming distance and popularity are
corresponding to $L=50$. The parameters in Heter-NBI and RE-NBI are
set as the ones corresponding to the lowest ranking scores (for
Heter-NBI \cite{Zhou2008}, $\beta_{\texttt{opt}}=-0.71$; for RE-NBI,
$a_{\texttt{opt}}=-0.75$). Each number presented in this table is
obtained by averaging over five runs, each of which has an
independently random division of training set and probe.}
\begin{center}
\begin{tabular} {ccccccc}
  \hline \hline
   Algorithms     & Ranking Score  &  Precision &  Intra-Similarity & Hamming Distance & Popularity  \\
   \hline
   GRM & 0.068 & 0.037 & 0.391 & 0.187 & 2612  \\
   UCF & 0.058 & 0.048 & 0.372 & 0.405 & 2381  \\
   OCF & 0.053 & 0.052 & 0.372 & 0.551 & 2065  \\
   NBI & 0.050 & 0.050 & 0.366 & 0.424 & 2366  \\
   Heter-NBI & 0.047 & 0.051 & 0.341 & 0.545 & 2197  \\
   RE-NBI & 0.039 & 0.062 & 0.336 & 0.629 & 2063  \\
   \hline \hline
    \end{tabular}
\end{center}
\end{table}

A recommendation algorithm should provide each user with an ordered
queue of all its uncollected objects. For an arbitrary target user
$u_i$, if the relation $u_i-o_j$ is in the probe set (accordingly,
in the training set, $o_j$ is an uncollected object for $u_i$), we
measure the position of $o_j$ in the ordered queue. For example, if
there are 1000 uncollected movies for $u_i$, and $o_j$ is the 10th
from the top, we say the position of $o_j$ is 10/1000, denoted by
$r_{ij}=0.01$. Since the probe entries are actually collected by
users, a good algorithm is expected to give high recommendations to
them, thus leading to small $r$. Therefore, the mean value of the
position value $\langle r\rangle$, called \emph{ranking score},
averaged over all the entries in the probe, can be used to evaluate
the algorithmic accuracy: the smaller the ranking score, the higher
the algorithmic accuracy, and vice verse. Note that, the number of
objects recommended to a user is often limited, and even given a
long recommendation list, the real users usually consider only the
top part of it. Therefore, we adopt in this paper another accuracy
index, namely \emph{precision}. For an arbitrary target user $u_i$,
the precision of $u_i$, $P_i(L)$, is defined as the ratio of the
number of $u_i$'s removed links (i.e., the objects collected by
$u_i$ in the probe), $R_i(L)$, contained in the top-$L$
recommendations to $L$, say
\begin{equation}
P_i(L)=R_i(L)/L.
\end{equation}
The precision of the whole system is the average of individual
precisions over all users, as:
\begin{equation}
P(L)=\frac{1}{m}\sum^m_{i=1}P_i(L).
\end{equation}
Since the ranking score does not depend on the length of
recommendation list, hereinafter, without a special statement, the
optimal value of a parameter always subjects to the lowest ranking
score. In Table 1 and Table 2, we report the algorithmic performance
for \emph{MovieLens} and \emph{Netflix}, respectively. If just
taking into account the recommendation accuracy, the network-based
inference performs better than global ranking method and
collaborative filtering (NBI performs remarkably better than UCF,
while better than OCF for ranking score and competitively to OCF for
precision).

\begin{figure}
\scalebox{1.5}[1.5]{\includegraphics{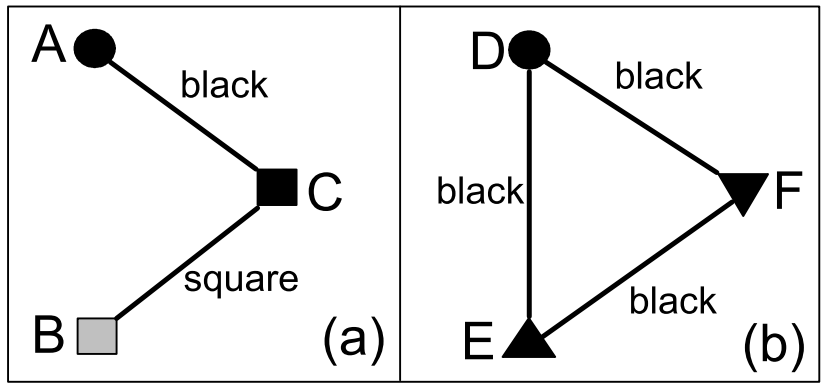}} \caption{Illustration
of redundant correlations.}
\end{figure}

\section{Improved Algorithm by Eliminating Redundant Correlations}
In NBI, for any user $u_i$, the recommendation value of an
uncollected object $o_j$ is contributed by all $u_i$'s collected
object, as
\begin{equation}
f_j'=\sum_lw_{jl}a_{li}.
\end{equation}
Those contributions, $w_{jl}a_{li}$, may result from the
similarities in same attributes, thus lead to heavy redundance. We
use an illustration, as shown in Fig. 2, to make our idea clearer.
Here, we assume that all the objects can be fully described by two
attributes, color and shape, and the target user, say $u_i$, likes
black and square. In Fig. 2(a), A and B are collected objects and C
is uncollected, while in Fig. 2(b), D and E are collected and F is
uncollected. All the five links, representing correlations between
objects, should have more or less the same weight in the
object-object network since each of them results from one common
attribute as labeled beside. Here the weight of each link is set to
be a unit.

\begin{figure}
\scalebox{1}[1]{\includegraphics{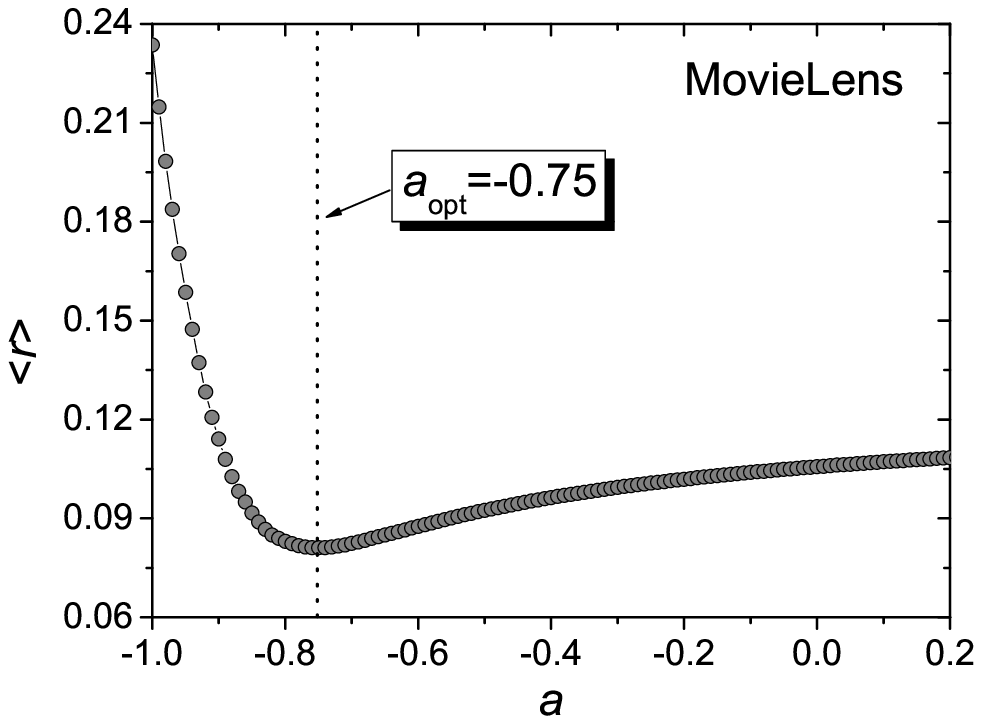}}
\scalebox{1}[1]{\includegraphics{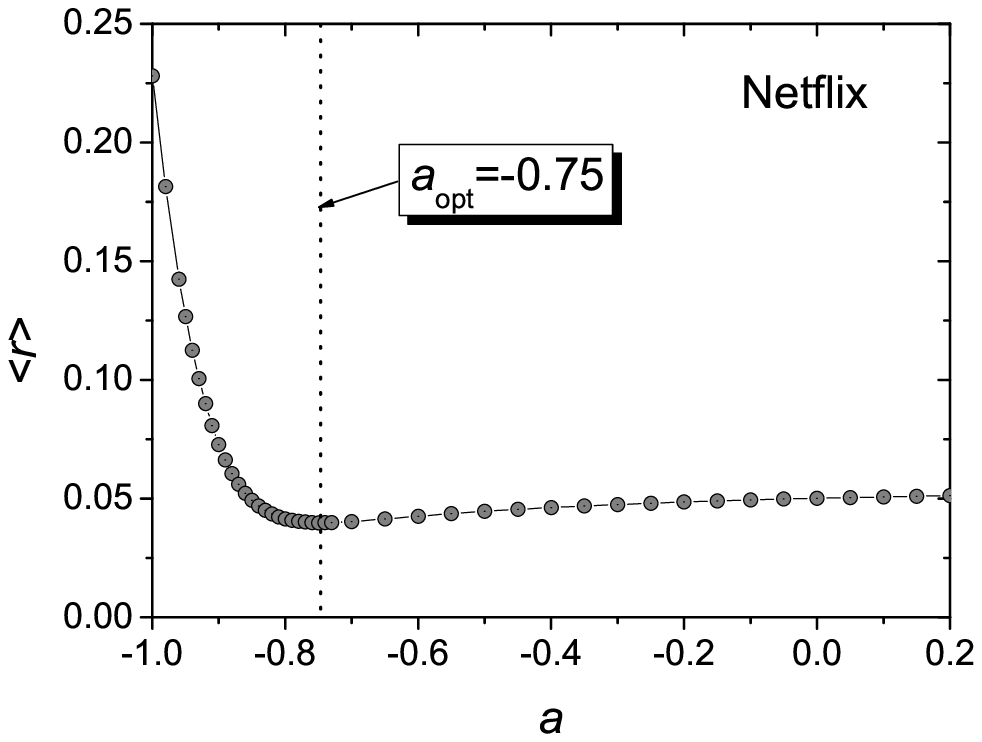}} \caption{The ranking
score $\langle r\rangle$ \emph{vs.} $a$. The upper and lower plots
shows the numerical results for \emph{MovieLens} and \emph{Netflix},
respectively. Each data point is obtained by averaging over five
runs, each of which has an independently random division of training
set and probe. Interestingly, for both \emph{MovieLens} and
\emph{Netflix}, the optimal $a$, corresponding to the minimal
$\langle r\rangle$, is $a_{\texttt{opt}} \approx -0.75$.}
\end{figure}

\begin{figure}
\scalebox{0.6}[0.6]{\includegraphics{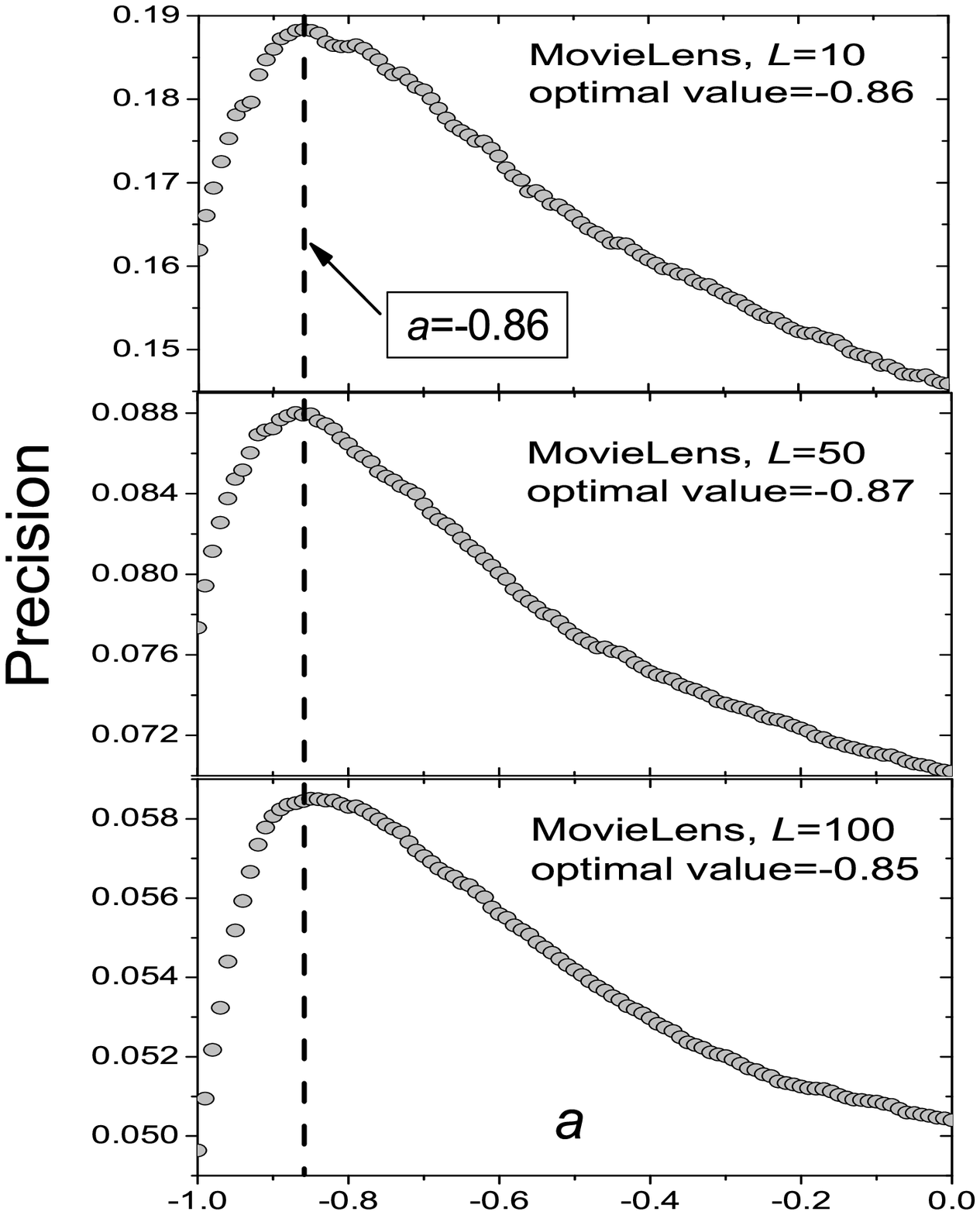}} \caption{The precision
vs. $a$ on \emph{MovieLens} data for some typical lengths of
recommendation list. Each data point is obtained by averaging over
five runs, each of which has an independently random division of
training set and probe.}
\end{figure}

\begin{figure}
\scalebox{0.6}[0.6]{\includegraphics{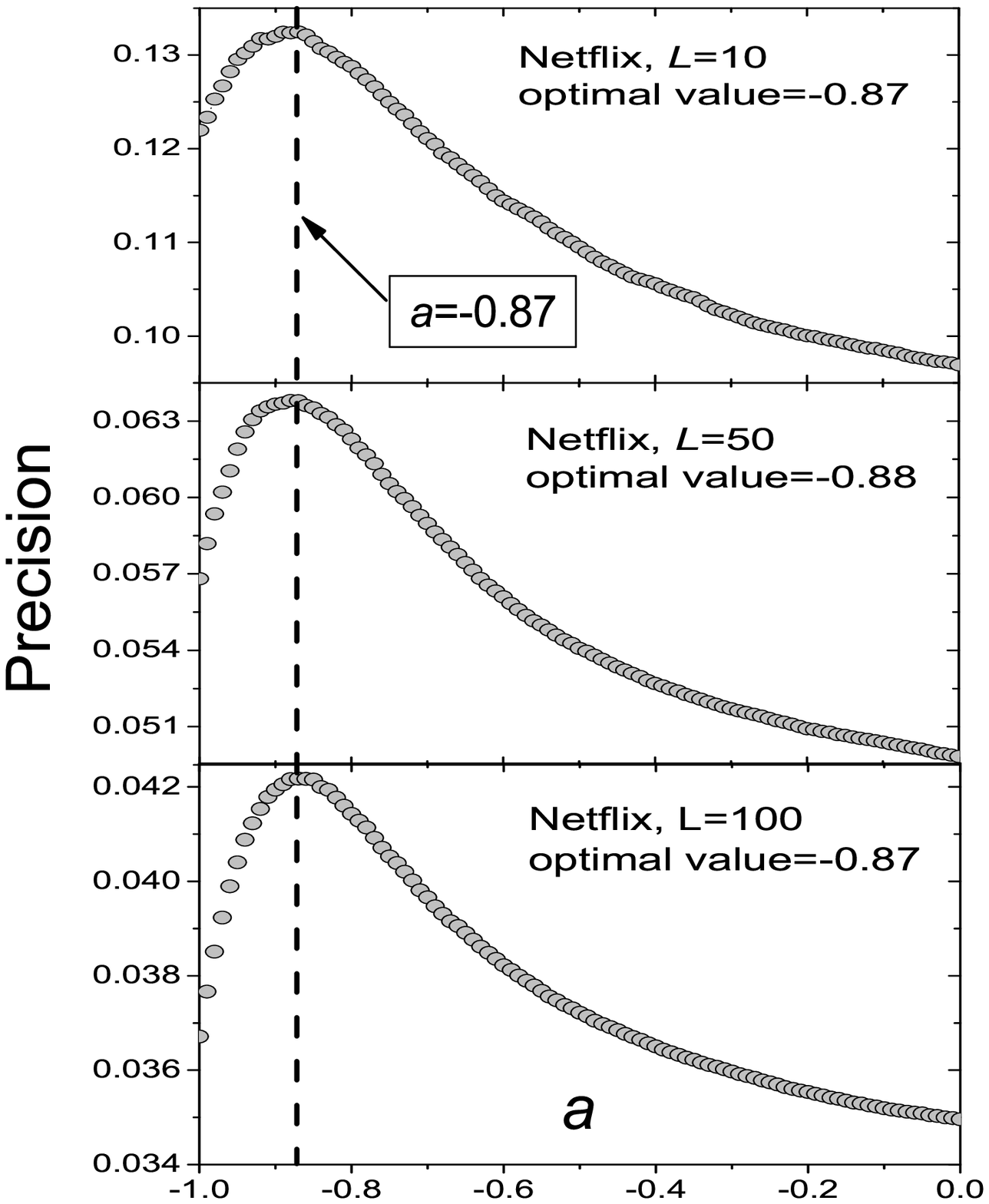}} \caption{The precision
vs. $a$ on \emph{Netflix} data for some typical lengths of
recommendation list. Each data point is obtained by averaging over
five runs, each of which has an independently random division of
training set and probe.}
\end{figure}

\begin{figure}
\scalebox{1.5}[1.5]{\includegraphics{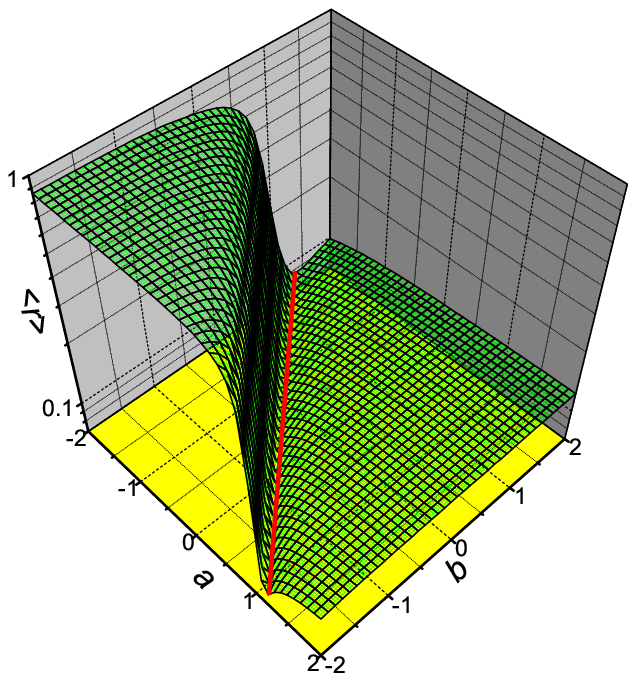}} \caption{The ranking
score $\langle r\rangle$ in $(a,b)$ plane for \emph{MovieLens} data.
The numerical simulations run over the parameters, $a$ and $b$, in
the interval $[-2,2]$ and $[-2,2]$ respectively, with step length
equaling 0.1. To make the figure clearer, the axis of $\langle
r\rangle$ is set to be logarithmic. Given $a$, denoting $b^*(a)$ the
optimal value of $b$ corresponding to the smallest $\langle
r\rangle$. The red thick line emphasizes approximately the function
$b^*(a)$. All the numerical results are obtained by averaging over
five independent runs with data division identical to the case shown
in Figure 3. The global minimum is $\langle r\rangle \approx
0.0794$, corresponding to $(a^*,b^*)=(-1.6,0.8)$.}
\end{figure}

For both C and F, the final recommendation value is two. However,
according to our assumption, the target user likes C more than F. It
is because in Fig. 2(a), the recommendations from A and B are
independent, resulted from two different attributes; while in Fig.
2(b), the recommendations resulting from the same attribute (i.e.,
color=black) are repeatedly counted twice. Indeed, when calculating
the recommendation value of F, the correlations D-F and E-F are
redundant for each other. Although the real recommendation systems
are much complicated than the simple example shown in Fig. 2, and no
clear classification of objects' attributes as well as no accurately
quantitative measurements of users' tastes can be extracted, we
believe the redundance of correlations is ubiquity in those systems,
which depresses the accuracy of NBI.

Note that, in Fig. 2(a), A and B, sharing no common property, do not
have any correlation (in real system, two objects, even without any
common/similar property, may have a certain weak correlation induced
by occasional collections). While in Fig. 2(b), D and E are tightly
connected for their common attribute, color=black, which is also the
very causer of redundant recommendations to F. Therefore, following
the path D$\rightarrow$E$\rightarrow$F, D and F have strong
second-order correlation. However, since the correlation between A
and B are very weak, the second-order correlation between A and C,
contributed by the path A$\rightarrow$B$\rightarrow$C, should be
neglectable.

Generally speaking, if the correlation between $o_i$ and $o_k$ and
the correlation between $o_j$ and $o_k$ contain some redundance to
each other, then the second-order correlation between $o_i$ and
$o_k$, as well as that between $o_j$ and $o_k$ should be strong.
Accordingly, subtracting the higher order correlations in an
appropriate way could, perhaps, further improve the algorithmic
accuracy. Motivated by this idea, we replace Eq. (5) by
\begin{equation}
\mathbf{f}'=(W+aW^2)\mathbf{f},
\end{equation}
where $a$ is a free parameter. When $a=0$, it degenerates to the
standard NBI discussed in the last section. If the present analysis
is reasonable, the algorithm with a certain negative $a$ could
outperforms the case with $a=0$.

Figure 3 reports the algorithmic accuracy, measured by the ranking
score, as a function of $a$, which has a clear minimum around
$a=-0.75$ for both \emph{MovieLens} and \emph{Netflix}. Compared
with the standard case (i.e., $a=0$), the ranking score can be
further reduced by 23\% for \emph{MovieLens} and 22\% for
\emph{Netflix}, respectively. This result strongly supports our
analysis. It is worthwhile to emphasize that, more than 20\% is
indeed a great improvement for recommendation algorithms. In
addition, we compare the present algorithm with the \emph{Latent
Dirichlet Allocation} (LDA) algorithm \cite{Blei2003}, which is
widely accepted as one of the most accurate personalized
recommendation algorithms thus far. Although LDA requires much more
computational time, the ranking score for \emph{MovieLens} data is
about 0.088, remarkably larger than the minimum, 0.082, obtained by
the present algorithm. The ultra accuracy of the present method,
even far beyond our expectation, indicates a great significance in
potential applications. In addition, in Fig. 4 and Fig. 5, we
present how the parameter $a$ affects the precision for some typical
lengths of recommendation list. Although the optimal value of $a$
leading to the highest precision is different from the one subject
to the lowest ranking score, the qualitative behaviors of $\langle
r\rangle$ versus $a$ and $P(L)$ versus $a$ are the same, that is, in
each case, there exists a certain negative $a$ corresponding to the
most accurate recommendations (subject to the specific accuracy
metric) with remarkable improvement compared with the standard NBI
at $a=0$. We compare the ranking score and precision for $L=50$ in
Table 1 and Table 2, where the Heter-NBI represents an improved NBI
algorithm with heterogenous initial resource distribution
\cite{Zhou2008}, and RE-NBI is the current algorithm. To be fair to
compare with parameter-free algorithms, in both Heter-NBI and
RE-NBI, the parameters are fixed as the ones corresponding to the
lowest ranking score, therefore the precisions presented in Table 1
and Table 2 are smaller than the optima. Even so, the present
algorithm give much more accurate recommendations than all others.

\begin{figure}
\scalebox{1.2}[1.2]{\includegraphics{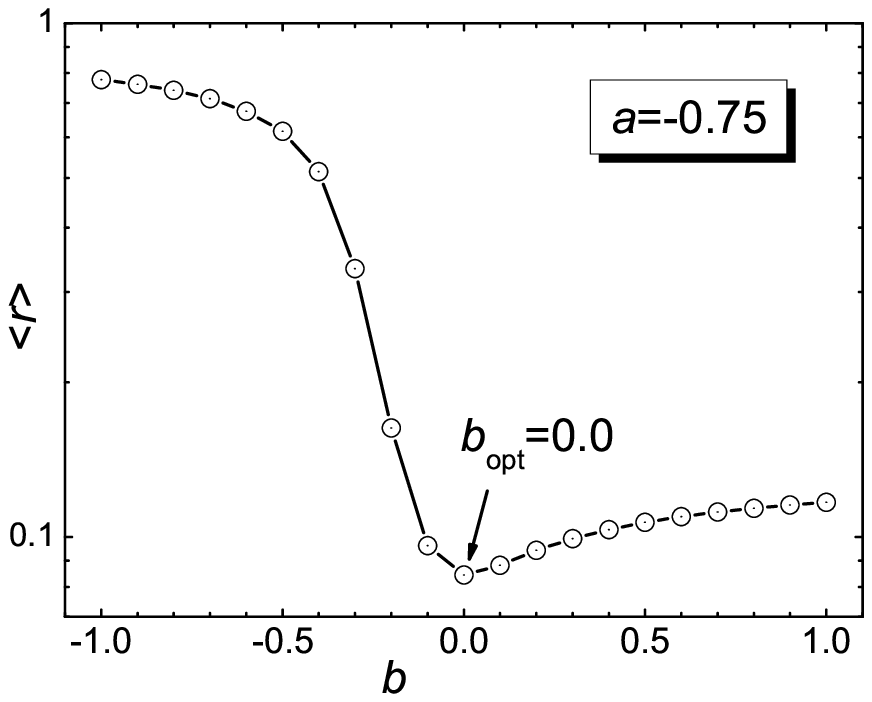}} \caption{The ranking
score $\langle r\rangle$ \emph{vs.} $b$ given $a=-0.75$ for
\emph{MovieLens} data. All the numerical results are obtained by
averaging over five independent runs with data division identical to
the case shown in Fig. 3. The optimal value of $b$ is zero.}
\end{figure}

Although without a clear physical picture, Eq. (13) can be naturally
extended to a formula containing even higher order of correlations
than $W^2$, such as
\begin{equation}
\mathbf{f}'=(W+aW^2+bW^3)\mathbf{f},
\end{equation}
where $b$ is also a free parameter. Since the computational
complexity increases quickly as the increasing of the highest order
of $W$, one should check very carefully if such kind of extension is
valuable.

Extensively numerical simulations have been done to search the
global minimum of $\langle r\rangle$ in $(a,b)$ plane for
\emph{MovieLens} data. Given $a$, denoting $b^*(a)$ the optimal
value of $b$ corresponding to the smallest $\langle r\rangle$, as
shown in Fig. 6 (red thick line), $b^*(a)$ decreases with the
increasing of $a$ in aa approximately linear way. The global minimum
of $\langle r\rangle$ is about 0.0794, corresponding to
$(a^*,b^*)=(-1.6,0.8)$. That is to say, taking into account the cube
of $W$, the algorithmic accuracy can be further improved by about
3\%. However, the readers should be warned that the optimal
parameters, $a^*$ and $b^*$, may be far different for different
systems, and finding out them will take very long time for huge size
systems. Therefore, the algorithm concerning three or even higher
order of weighted matrix may be not applicable in real systems.

\begin{figure}
\scalebox{1}[1]{\includegraphics{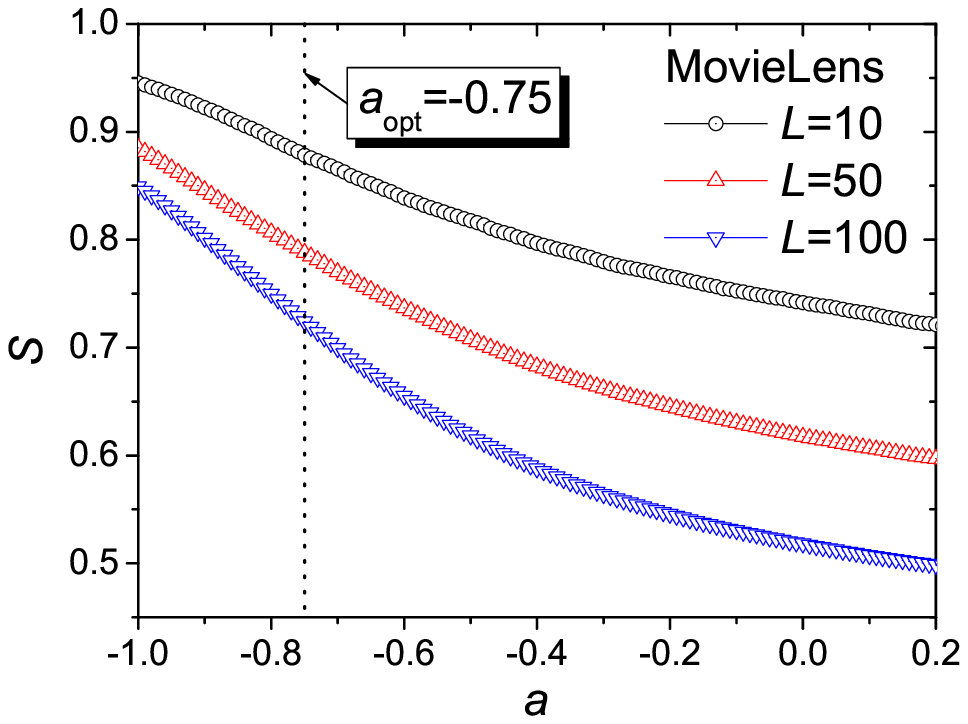}}
\scalebox{1}[1]{\includegraphics{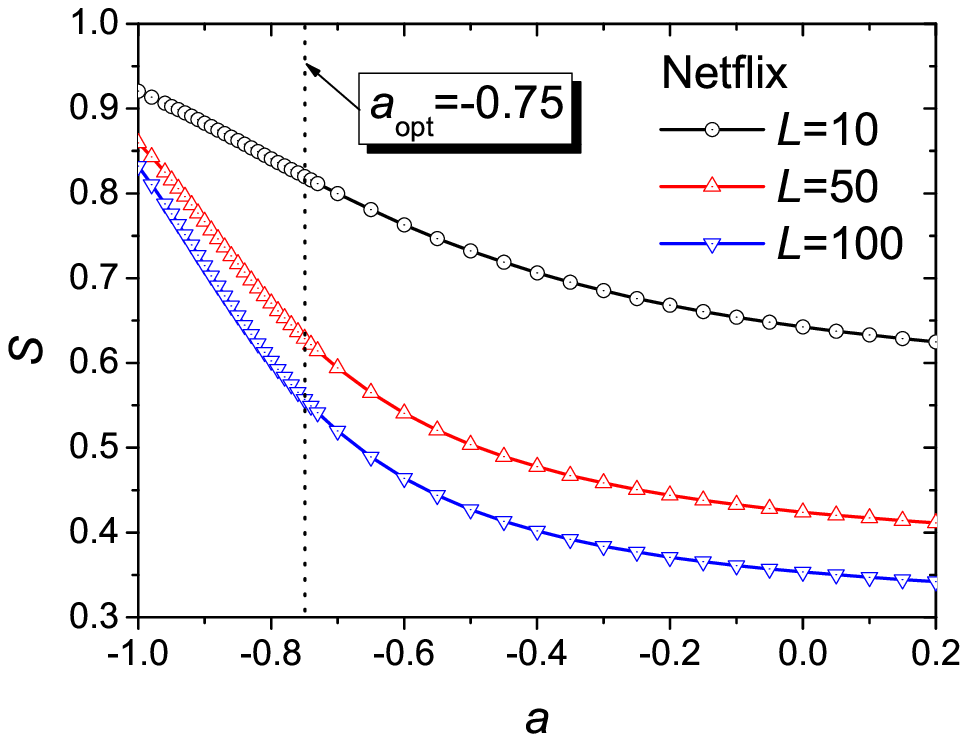}} \caption{The Hamming
distance, $S$, as a function of $a$. The black circles, red
up-triangles and blue down-triangles represent the cases with
typical lengths $L=10$, 50 and 100, respectively. The upper and
lower plots correspond to the results on \emph{MovieLens} and
\emph{Netflix}, respectively. The vertical line marks the optimal
value of $a$, as $a_{\texttt{opt}}=-0.75$. All the numerical results
are obtained by averaging over five independent runs with data
division identical to the case shown in Figure 3.}
\end{figure}

Instead of the global search in $(a,b)$ plane, a possible way to
quickly find out a nearly minimal $\langle r\rangle$ is using a
greedy algorithm containing two steps. First, we search the optimal
$a$ considering only the square of $W$, as shown in Eq. (13). Then,
we search the optimal $b$ with $a$ fixed as the optimal value
obtained in the first step. Clearly, this greedy method runs much
faster than the blinding search in $(a,b)$ plane. However, as shown
in Fig. 7, for \emph{MovieLens} data with $a_{\texttt{opt}}=-0.75$,
the optimal $b$ is zero, giving no improvement of the algorithm
shown in Eq. (13). Therefore, though the introduction of two order
correlation can greatly improve the algorithmic accuracy, to
consider three or even higher order of $W$ may be not valuable.

\section{Popularity and Diversity of Recommendations}
When judging the algorithmic performance, most of the previous works
only consider the accuracy of recommendations. Those measurements
include \cite{Adomavicius2005,Herlocker2004,Zhou2007,Huang2004}
\emph{ranking score}, \emph{hitting rate}, \emph{precision},
\emph{recall}, \emph{F-measure}, and so on. However, besides
accuracy, two significant ingredients must be taken into account.
Firstly, the algorithm should guarantee the diversity of
recommendations, viz., different users should be recommended
different objects. It is also the soul of personalized
recommendations. The inter-diversity can be quantified via the
\emph{Hamming distance} \cite{Zhou2008}. Denoting $L$ the length of
recommendation list (i.e., the number of objects recommended to each
user), if the overlapped number of objects in $u_i$ and $u_j$'s
recommendation lists is $Q$, their Hamming distance is defined as
\begin{equation}
H_{ij}=1-Q/L.
\end{equation}
Generally speaking, a more personalized recommendation list should
have larger Hamming distances to other lists. Accordingly, we use
the mean value of Hamming distance,
\begin{equation}
S=\frac{1}{m(m-1)}\sum_{i\neq j} H_{ij},
\end{equation}
averaged over all the user-user pairs, to measure the diversity of
recommendations. Note that, $S$ only takes into account the
diversity among users. Besides, a good algorithm should also make
the recommendations to a single user diverse to some extent
\cite{Ziegler2005}, otherwise users may feel tired for receiving
many recommended objects under the same topic. Motivated by Ziegler
\emph{et al.} \cite{Ziegler2005}, for an arbitrary target user
$u_l$, denoting the recommended objects for $u_l$ as
$\{o_1,o_2,\cdots,o_L\}$, the \emph{intra-similarity} of $u_l$'s
recommendation list can be defined as:
\begin{equation}
I_l=\frac{1}{L(L-1)}\sum_{i\neq j} s^o_{ij},
\end{equation}
where $s^o_{ij}$ is the similarity between objects $o_i$ and $o_j$,
as shown in Eq. (8). The intra-similarity of the whole system is
thus defined as:
\begin{equation}
I=\frac{1}{m}\sum^m_{l=1} I_l.
\end{equation}
In this paper, we use $S$ and $I$ respectively quantify the
diversities among recommendation lists and inside a recommendation
list.

Secondly, with more or less the same accuracy, an algorithm that
recommends less popular objects is better than the one recommending
popular objects. Taking recommender systems for movies as an
example, since there are countless channels to obtain information of
popular movies (TV, the Internet, newspapers, radio, etc.),
uncovering very specific preference, corresponding to unpopular
objects, is much more significant than simply picking out what a
user likes from the top-viewed movies. The popularity can be
directly measured by the average degree $\langle k\rangle$ over all
the recommended objects.

\begin{figure}
\scalebox{1}[1]{\includegraphics{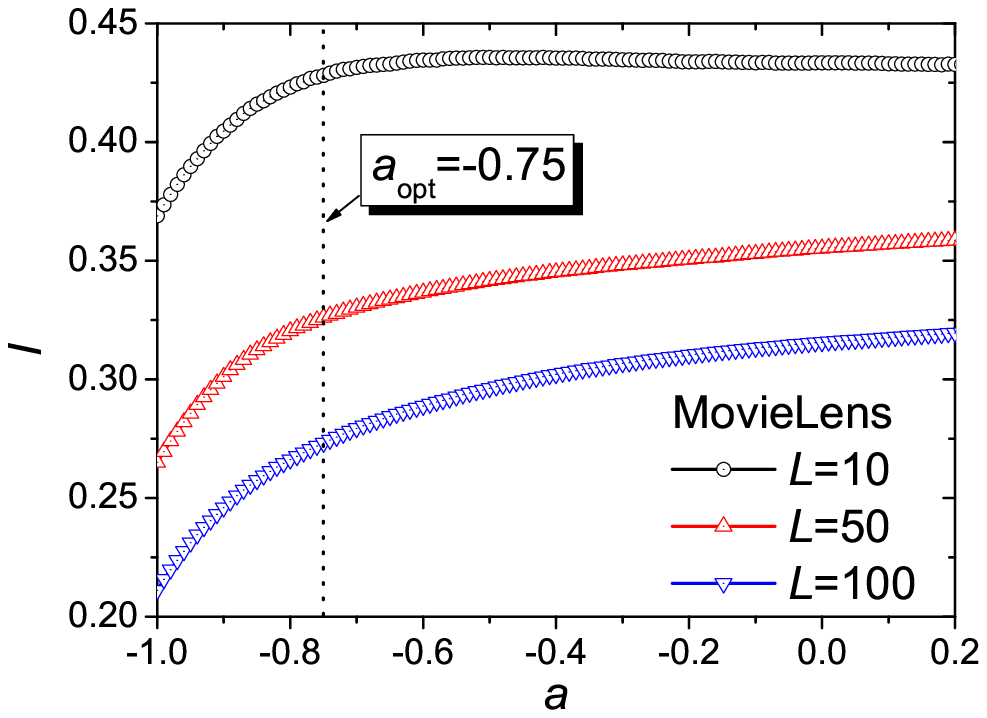}}
\scalebox{1}[1]{\includegraphics{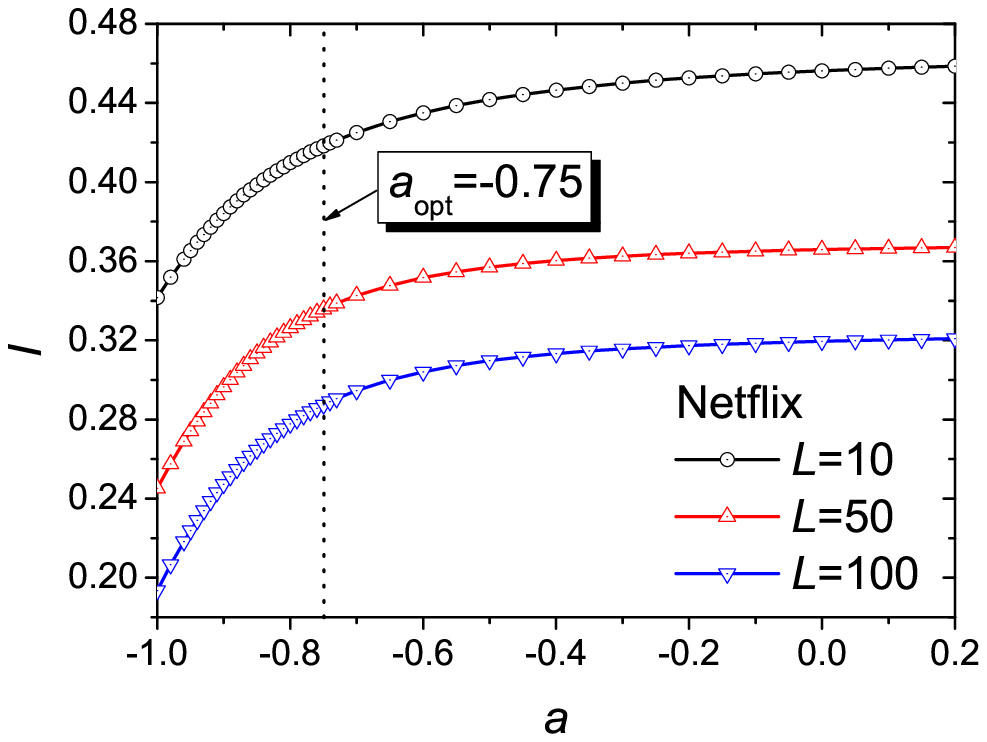}} \caption{The
intra-similarity, $I$, as a function of $a$. The black circles, red
up-triangles and blue down-triangles represent the cases with
typical lengths $L=10$, 50 and 100, respectively. The upper and
lower plots correspond to the results on \emph{MovieLens} and
\emph{Netflix}, respectively. The vertical line marks the optimal
value of $a$, as $a_{\texttt{opt}}=-0.75$. All the numerical results
are obtained by averaging over five independent runs with data
division identical to the case shown in Figure 3.}
\end{figure}

Statistically speaking, the recommendations displaying high
inter-diversity (i.e., large $S$) will have small popularity. It is
because those high-degree objects (i.e., popular objects) are always
the minority in a real system, and highly diverse recommendation
lists must involve many less popular objects, thus depress the
average degree $\langle k\rangle$. In contrast, a smaller $\langle
k\rangle$ does not guarantee a higher $S$. An extreme example is to
recommend every user the uncollected objects with minimal degrees.
Therefore the average degree reaches its minimum, while the Hamming
distance is close to zero since the recommendations to every user
are almost the same. Therefore, $S$ of recommendations provides more
information for the algorithmic performance than $\langle k\rangle$.
However, the calculation of $S$ takes much longer time than that of
$\langle k\rangle$, especially for a system containing quite a
number of users. In addition, the definition of popularity is
simpler and more intuitional than Hamming distance. In comparison,
the intra-similarity, $I$, mainly concerning the underlying content
of objects (two objects with similar content or in the same category
usually have high probability to be collected by same users), is not
directly relevant to the popularity. Therefore, we user all the
three metrics here to provide comprehensive evaluation. In a word,
besides the accuracy, an algorithm giving higher $S$, lower $I$ and
lower $\langle k\rangle$ is more favorable.

\begin{figure}
\scalebox{1}[1]{\includegraphics{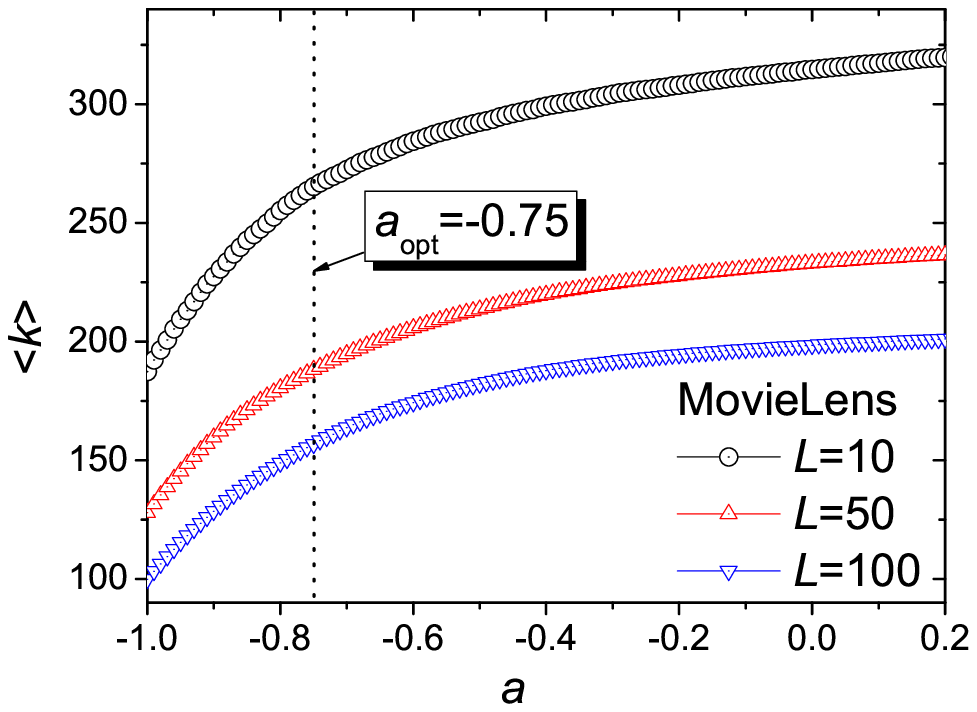}}
\scalebox{1}[1]{\includegraphics{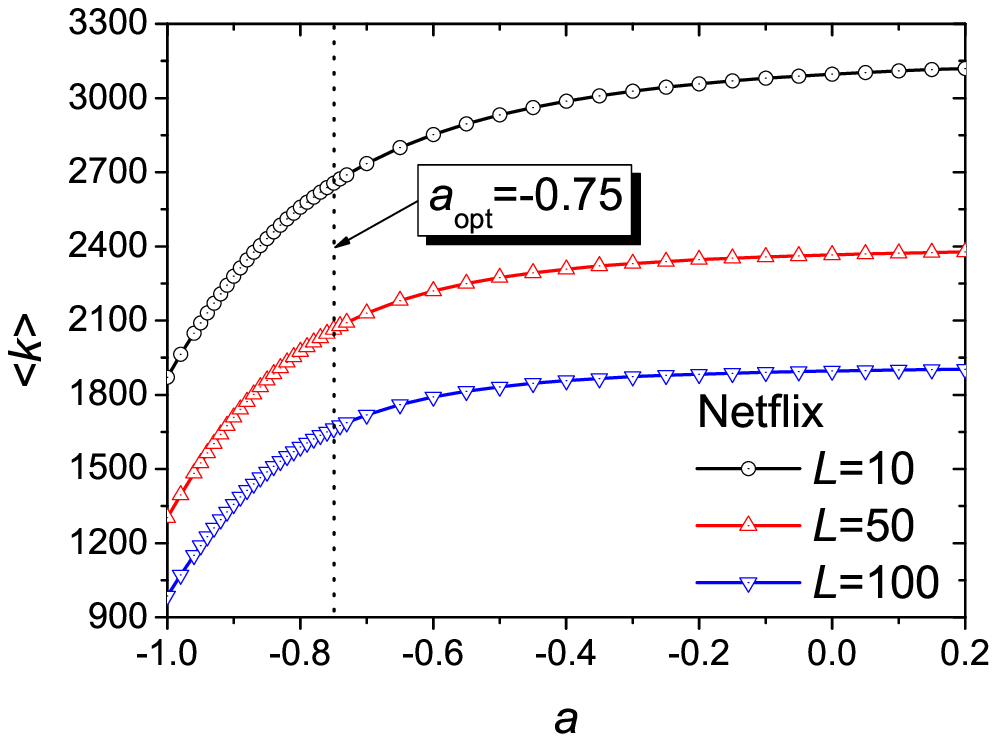}} \caption{The average
degree, $\langle k\rangle$, as a function of $a$. The black circles,
red up-triangles and blue down-triangles represent the cases with
typical lengths $L=10$, 50 and 100, respectively. The upper and
lower plots correspond to the results on \emph{MovieLens} and
\emph{Netflix}, respectively. The vertical line marks the optimal
value of $a$, as $a_{\texttt{opt}}=-0.75$. All the numerical results
are obtained by averaging over five independent runs with data
division identical to the case shown in Figure 3.}
\end{figure}

In Fig. 8, we report the numerical results about how the parameter
$a$ affects the Hamming distance, $S$. From this figure, one can see
that the behaviors of $S(a)$ for both \emph{MovieLens} and
\emph{Netflix}, as well as for different $L$, are qualitatively the
same, namely $S$ is negatively correlated with $a$: the smaller $a$
the higher $S$. As a result, the present algorithm with $a=-0.75$
can provide obviously higher inter-diverse recommendations compared
with the standard NBI at $a=0$. Figure 9 and Figure 10 show how the
parameter $a$ affects the intra-similarity $I$ and the popularity
$\langle k\rangle$, respectively. Clearly, the smaller $a$ leads to
less intra-similarity and popularity, and thus the present algorithm
can find its advantage in recommending less popular objects with
diverse topics to users, compared with the standard NBI. Generally
speaking, the popular objects must have some attributes fitting the
tastes of the masses of the people. The standard NBI may repeatedly
count those attributes and thus give overstrong recommendation for
the popular objects, which increases the average degree of
recommendations, as well as reduces the diversity. The collaborative
filtering, considering only the first-order correlations, has the
same problem as the standard NBI. The present algorithm with
negative $a$ can to some extent eliminate the redundant
correlations, namely assigns lower weights to the everyone-like
attributes, and thus give higher chances to less popular objects and
the objects with diverse topics different from the mainstream.

We summary the algorithmic performance in Table 1 and Table 2. One
can find that in the case $a=-0.75$, the present algorithm
outperforms the standard network-based inference (NBI) (i.e., $a=0$)
\cite{Zhou2007} and its variant with heterogenous initial resource
distribution (Heter-NBI) \cite{Zhou2008} in all five criteria: lower
ranking score, higher precision, larger Hamming distance, lower
intra-similarity and smaller average degree.

\section{Conclusion and Discussion}
The network-based inference \cite{Zhou2007}, as introduced in
Section 2, has higher accuracy as well as lower computational
complexity than the widest personalized recommendation algorithm,
namely the user-based collaborative filtering. Therefore, it has
great potential significance in practical purpose. However, in this
paper, we point out that in the NBI, the correlations resulting from
a specific attribute may be repeatedly counted in the cumulative
recommendations from different objects. Those redundant correlations
will depress the algorithmic accuracy. By considering the higher
order correlations, $W^2$, we design an effective algorithm that
can, to some extent, eliminate the redundant correlations. The
algorithmic accuracy, measured by the ranking score, can be further
improved by 23\% for \emph{MovieLens} data and 22\% for
\emph{Netflix} data in the optimal case at $a=-0.75$. Since the
algorithm considering even higher order of $W$ takes too long time
to be applied in real systems, and the improvement is not much as
shown in Fig. 6 and Fig. 7, we suggest the readers taking into
account $W$ and $W^2$ only. The current method can also be naturally
extended to deal with the multi-rating recommender systems (see Ref.
\cite{Zhang2007b} a diffusion-like algorithm for multi-rating
recommender systems). Indeed, the numerical simulations on the
\emph{Netflix} data with ratings show that considering the
second-order correlations in building the transfer matrix can also
improve the prediction accuracy (the method are more complicated
than the binary systems present in this paper, see Ref.
\cite{Zhang2007b} for details).

Most of the previous studies considered the algorithmic accuracy
only. Here, we argue that the diversity and popularity, as the
significant criteria of algorithmic performance, should also be
taken into account. Diversity is the soul of a personalized
recommendation algorithm, that is to say, different users should be
recommended, in general, different objects, and for a single user,
the objects recommended to him should contain diverse topics. In
addition, the recommendations of less popular objects are very
significant in the modern information era, since those objects, even
perfectly match a user's tastes, could never be found out by this
user himself from countless congeneric objects (e.g., millions of
books and billions of webs). Without recommendation algorithm, those
very less popular objects look like the dark information for normal
users. Therefore, the algorithm that can provide accurate
recommendations for less popular objects can be considered as a
powerful tool uncovering the dark information. In a word, with more
or less the same accuracy, an algorithm giving higher diversity and
lower popularity is more favorable, and the numerical results show
that the present algorithm can outperform the standard NBI and both
the user-based and object-based collaborative filtering algorithms
simultaneously in all five criteria: lower ranking score, higher
precision, larger Hamming distance, lower intra-similarity and
smaller average degree.

How to better provide personalized recommendations is a
long-standing challenge in modern information science. Any answer to
this question may intensively change our society, economic and life
style in the near future. We believe the current work can enlighten
readers in this interesting and exciting direction.

\section*{Acknowledgments}
We acknowledge \emph{GroupLens Research Group} for \emph{MovieLens}
data. This work is benefitted from Matus Medo who has tested the
present method (an extended version) in a multi-rating
recommendation system (based on the \emph{Netflix} data) and also
found a certain improvement, and Cihang Jin who has provided us the
ranking score on \emph{MovieLens} data by using the Latent Dirichlet
Allocation (LDA) algorithm. This work is partially supported by SBF
(Switzerland) for financial support through project C05.0148
(Physics of Risk), the Swiss National Science Foundation
(205120-113842), and the Future and Emerging Technologies (FET)
programme within the Seventh Framework Programme for Research of the
European Commission, under FET-Open grant number 213360 (LIQUIDPUB
project). T.Z. and B.H.W. acknowledge the National Natural Science
Foundation of China under Grant Nos. 10635040 and 60744003, as well
as the 973 Project 2006CB705500.

\end{document}